\documentclass[twocolumn,showpacs,preprintnumbers,amsmath,amssymb]{revtex4}
\usepackage{amssymb}
\usepackage{amsmath}
\usepackage{graphicx}
\begin{document}

\title{\bf Doping dependance of the spin resonance peak in bilayer
high-$T_c$ superconductors}

\author
{
 T. Zhou$^{1}$, Z. D. Wang$^{1,2}$, and Jian-Xin Li$^{2}$
}

\affiliation{ $^{1}$Department of Physics and Center of Theoretical
and Computational Physics, University of Hong Kong, Pokfulam
Road, Hong Kong, China\\
$^{2}$National Laboratory of Solid State Microstructures and
Department of Physics, Nanjing University, Nanjing 210093, China\\
}

\date{\today}

\begin{abstract}
Motivated by a recent experiment on the bilayer
Y$_{1-x}$Ca$_{x}$Ba$_2$Cu$_3$O$_y$ superconductor and based on a
bilayer $t-J$ model, we calculate the spin susceptibility at
different doping densities in the even and odd channels in a bilayer
system. It is found that the intensity of the resonance peak in the
even channel is much weaker than that in the odd one, with the
resonance position being at a higher frequency. While this
difference decreases as the doping increases, and both the position
and amplitude of the resonance peaks in the two channels are very
similar in the deeply overdoped sample. Moreover, the resonance
frequency in the odd channel is found to be linear with the critical
temperature $T_c$, while the resonance frequency increases as doping
decreases in the even channel  and tends to saturate at the
underdoped sample. We elaborate the results based on the Fermi
surface topology and the $d$-wave superconductivity.

\end{abstract}

 \maketitle

\section{introduction}
Inelastic neutron scattering (INS) experiments have been playing an
important role in the studies of the spin dynamics of high-$T_c$
superconductors. They can provide direct information of the momentum
and frequency dependence of the dynamical spin susceptibility. Over
the past decade, one of the most striking features observed in the
INS experiments is the resonant spin excitation. The resonance peak,
which has been found in several classes of cuprate
materials~\cite{ross,fong,heh,roso,step}, has attracted much
experimental and theoretical attention. This peak is centered at the
momentum $(\pi,\pi)$, with its intensity decreasing rapidly when the
frequency moves away from $(\pi,\pi)$. The resonance frequency is
found to be in proportional to the critical
temperature~\cite{roso,step,dai}. Theoretically, the origin of the
spin resonance and its role on superconductivity are still open
questions~\cite{bat,sei,and,lin,seg,erem,morr}. It has been proposed
that the spin resonance is a collective spin excitation
mode~\cite{morr,bri,norman,jxli,kao,li,jxl,zhou}. Based on this
scenario, many properties of spin fluctuations observed in the INS
experiments have be explained consistently.

In the YBa$_2$Cu$_3$O$_y$ (YBCO) and Bi$_2$Sr$_2$CaCu$_2$O$_{8+x}$
(Bi-2212) family, one unit cell contains two CuO$_2$ planes. The
electronic states in different CuO$_2$ layers belonging to one cell
are strongly coupled at all doping levels. Thus, two modes of
magnetic excitation are expected to exist in the bilayer systems,
i.e., one in the even channel and the other in the odd channel
according to the symmetry with respect to the exchange of the
layers~\cite{mil,taoli,litao,esc}. In earlier experiments, this
expectation was only confirmed in the insulating YBCO
samples~\cite{rezn}. In the superconducting state, the spin
resonance mode was not observed in the even channel, presumably due
to a much weaker intensity in this channel. Recently, the
instrumentation advances have made it possible to resolve weaker
features in the INS experiments. Two distinct resonance modes were
observed in the superconducting state of bilayer
(Y,Ca)Ba$_2$Cu$_3$O$_y$ samples ~\cite{pail,pai}. It was found that
the resonance peak intensity in the even channel ($I^e$) is much
weaker than that in the odd channel ($I^o$), and the resonance
frequency is higher than that of the odd channel. Very recently, the
doping evolution of the resonance peak in both the even and odd
channels of (Y,Ca)Ba$_2$Cu$_3$O$_y$ was studied in detail by the INS
experiments~\cite{pailh}. In the overdoped samples $(y=7)$, the
resonance position of the odd channel is close to that of the even
channel. At this doping level, the two resonance modes have also
closer intensities ($I^e/I^o=0.4$). When the doping density
decreases, the doping evolution of the resonance frequency in the
odd channel seems to follow a similar doping dependence as $T_c$,
while the resonance frequency seems to keep increasing in the even
channel as the doping decreases and saturates to a constant in the
underdoped sample. Moreover, the resonance peak intensity in the odd
channel is also much larger than that in the even channel in the
underdoped sample. The intensity ratio $I^e/I^o$ decreases
monotonously as doping decreases and reaches 0.05 in strongly
underdoped samples.

Motivated by these experimental observations, we here present a
detailed investigation of the doping dependence of the spin
resonance mode in the even and odd channels. Following
Ref.~\cite{lijx}, we employ a bilayer $t-J$ type Hamiltonian
including the interlayer hopping and interlayer exchange coupling.
In order to examine the robustness of the doping dependance of the
difference between the two resonance modes, we also look into in
detail the effect of the interlayer hopping parameters $t_\perp$ and
the interlayer exchange coupling $J_\perp$.

The article is organized as follows. In Sec. II, we introduce the
model and work out the formalism. In Sec. III, we perform numerical
calculations and discuss the obtained results. Finally, we give a
brief summary in Sec. IV.

\section{Hamiltonian and Formalism}
We start with a Hamiltonian which describes a system with two layers
per unit cell.

\begin{eqnarray}
H&=&-t\sum_{\langle ij\rangle l}
c^{(l)\dagger}_{i\sigma}c^{(l)}_{j\sigma}-h.c.-t^{\prime}\sum_{\langle
ij\rangle^{\prime} l}
c^{(l)\dagger}_{i\sigma}c^{(l)}_{j\sigma}-h.c.\nonumber\\&&-\widetilde{t}_\perp\sum_{ij}
c^{(1)\dagger}_{i\sigma}c^{(2)}_{j\sigma}-h.c. +J\sum_{\langle
ij\rangle l}S^{(l)}_i\cdot S^{(l)}_j\nonumber\\&&+J_\perp\sum_{ij}
S^{(1)}_i\cdot S^{(2)}_j,
\end{eqnarray}
where $l=1,2$ denotes the layer index. In the slave-boson approach,
the physical electron operators $c^{(l)}_{i\sigma}$ are expressed by
slave bosons $b^{(l)}_i$ carrying the charge and fermions
$f^{(l)}_{i\sigma}$ representing the spin,
$c^{(l)}_{i\sigma}=b^{(l)\dagger}_if^{(l)}_{i\sigma}$. At the
mean-field level, we consider the order parameters
$\Delta^{(l)}_{ij}=\langle
f^{(l)}_{i\uparrow}f^{(l)}_{j\downarrow}-f^{(l)}_{i\downarrow}f^{(l)}_{j\uparrow}\rangle=\pm\Delta_0$,
($\pm$ depend on if the bond $\langle ij\rangle$ is in the $\hat{x}$
or the $\hat{y}$ direction), $\chi^{(l)}_{ij}=\sum_{\sigma}\langle
f^{(l)\dagger}_{i\sigma}f^{(l)}_{j\sigma}\rangle=\chi_{0}$. In the
superconducting state, bosons condense
$b^{(l)}_i\rightarrow<b^{(l)}_i>=\sqrt{\delta}$, where$\delta$ is
the hole concentration.

Then, the mean-field Hamiltonian can be written as,
\begin{eqnarray}
H_m&=&\sum_{{\bf k}\sigma l}\varepsilon_{\bf k}f^{(l)\dagger}_{{\bf
k}\sigma}f^{(l)}_{{\bf k}\sigma} -\sum_{{\bf k}l}\Delta_{\bf
k}(f^{(l)\dagger}_{{\bf k}\uparrow}f^{(l)\dagger}_{-{\bf
k}\downarrow}+h.c.)\nonumber\\&& +\sum_{{\bf k}\sigma}[t_\perp
e^{ik_zc}f^{(1)\dagger}_{{\bf k}\sigma}f^{(2)}_{{\bf
k}\sigma}+h.c.]+\varepsilon_0,
\end{eqnarray}
with $\varepsilon_{\bf k}=-2(\delta t+J'\chi_0)(\cos k_x+\cos
k_y)-4\delta t'\cos k_x \cos k_y-\mu$, $\Delta_{\bf
k}=2J'\Delta_0(\cos k_x-\cos k_y)$,
$\varepsilon_0=4NJ'(\chi^{2}_0+\Delta^{2}_0)$, and
$J^{\prime}=3J/8$. Diagonalizing the Hamiltonian, we can get the
antibonding band (A) and bonding band (B) with the dispersion
$\xi^{(A,B)}_{\bf k}=\varepsilon_{\bf k}\pm t_\perp$. Here we use
the momentum independent interlayer hopping constant $t_\perp$,
being consistent with the recent angle resolved photoemission
experiment on YBCO~\cite{bori}, which reveals an obvious bilayer
splitting along the nodal direction.

The bare spin susceptibility can be expressed as,
{\setlength\arraycolsep{-4pt}
\begin{eqnarray}
\lefteqn{ \chi^{(\alpha,\alpha^{\prime})}({\bf q},\omega)=
\frac{1}{N}\sum_{\bf k}\{\frac{1}{4} (1-\frac{\xi_{\bf
k}^{(\alpha)}\xi_{{\bf k}+{\bf q}}^{(\alpha^{\prime})}+\Delta_{\bf
k}\Delta_{{\bf k}+{\bf q}}}{E_{\bf k}^{(\alpha)}E_{{\bf k}+{\bf
q}}^{(\alpha^{\prime})}}) }\nonumber\\&&[\frac{1-f(E_{{\bf k}+{\bf
q}}^{(\alpha^{\prime})})-f(E_{\bf k}^{(\alpha)})}{\omega+(E_{{\bf
k}+{\bf q}}^{(\alpha^{\prime})}+E_{\bf k}^{(\alpha)})+i\Gamma}-
\frac{1-f(E_{{\bf k}+{\bf q}}^{(\alpha^{\prime})})-f(E_{\bf
k}^{(\alpha)})}{\omega-(E_{{\bf k}+{\bf
q}}^{(\alpha^{\prime})}+E_{\bf k}^{(\alpha)})+i\Gamma}]\nonumber\\&&
+\frac{1}{2}[1+\frac{\xi_{\bf k}^{(\alpha)}\xi_{{\bf k}+{\bf
q}}^{(\alpha^{\prime})}+\Delta_{\bf k}\Delta_{{\bf k}+{\bf
q}}}{E_{\bf k}^{(\alpha)}E_{{\bf k}+{\bf q}}^{(\alpha^{\prime})}}]
\frac{f(E_{{\bf k}+{\bf q}}^{(\alpha^{\prime})})-f(E_{\bf
k}^{(\alpha)})}{\omega-(E_{{\bf k}+{\bf
q}}^{(\alpha^{\prime})}-E_{\bf k}^{(\alpha)})+i\Gamma}\}.
\end{eqnarray}}
Here $\alpha,\alpha^{\prime}=A,B$, $E_{\bf
k}^{(\alpha)}=\sqrt{\xi_{\bf k}^{(\alpha)2}+\Delta_{\bf k}^{2}}$ is
the quasiparticle energy, and $f(\omega)$ is the Fermi distribution
function.

The bare even and odd channel spin susceptibilities which come
respectively from the intraband and interband electronic
transitions, are given by
\begin{eqnarray}
\chi_0^{e}({\bf q},\omega)&=&\chi^{(A,A)}({\bf
q},\omega)+\chi^{(B,B)}({\bf q},\omega),\nonumber\\
\chi_0^{o}({\bf q},\omega)&=&\chi^{(A,B)}({\bf
q},\omega)+\chi^{(B,A)}({\bf q},\omega).
\end{eqnarray}

By including the correction of the antiferromagnetic (AF) spin
fluctuations to the spin susceptibility in the form of the
random-phase approximation (RPA), the renormalized spin
susceptibilities for the even and odd channels can be obtained as
\begin{equation}
\chi^{e(o)}({\bf q},\omega)=\frac{\chi^{e(o)}_0({\bf
q},\omega)}{1+(\widetilde{\alpha} J_{\bf q}\pm J_\perp)\times
\chi^{e(o)}_0({\bf q},\omega)/2},
\end{equation}
where $\pm$ signs represent the spin susceptibility in the even and
odd channels, respectively, $J_\perp$ is the interlayer exchange
coupling, $J_{\bf q}=J(\cos q_x+\cos q_y)$ is the intralayer
exchange. We here also include $\widetilde{\alpha}$ to set the AF
instability at $\delta=0.02$~\cite{bri}. The mean-field order
parameters $\chi_0$, $\Delta_0$ together with the chemical potential
$\mu$ for different doping $\delta$ can be obtained from the
self-consistent equations. The other parameters we choose are
$t=2J$, $t^{\prime}=-0.45t$.

\begin{figure}

\centering

\includegraphics[scale=0.4]{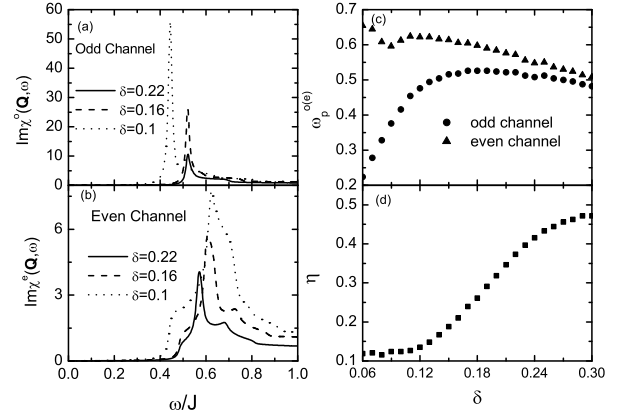}
\caption{Panels (a) and (b) are the imaginary parts of the spin
susceptibility versus the frequency for the wave vector ${\bf
Q}=(\pi,\pi)$ with $t_\perp=0.1J$, $J_\perp=0.15J$ in the odd and
even channels, respectively. Panel (c) is the resonance position as
a function of the doping. Panel (d) is the intensity ratio of the
spin susceptibilities between the even and odd channels. The
quasiparticle damping $\Gamma=0.01$.}\label{fig1}
\end{figure}

Before we present our results, we wish to point out that the above
formulas represent the spin susceptibility of the fermions. The spin
susceptibility for physical electrons should be $\delta^{2}\chi$ due
to the boson condensation in the superconducting state.

\section{results and discussion}

\begin{figure}

\centering

\includegraphics[scale=0.6]{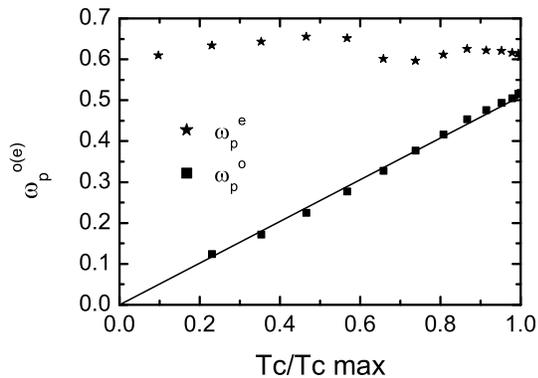}
\caption{The resonance frequency as a function of the critical
temperature $T_c$ with $t_\perp=0.1J$ and
$J_\perp=0.15J$.}\label{fig2}
\end{figure}

The odd and even channel spin susceptibilities at different doping
densities are shown in Figs.1(a) and (b), respectively. As seen, the
peak intensity in the odd channel is sensitive to the doping,
namely, it increases dramatically as the doping decreases. While in
the even channel it increases slowly as the doping decreases. To see
the doping evolution more clearly, we plot the peak positions of the
even and odd channels as a function of the doping in Fig.1(c), and
the ratio of the intensities of the spin resonance between the even
and odd channels, $\eta=I^{e}/I^{o}$ versus doping in Fig.1(d). From
Fig.1(c), the resonance frequency in the odd channel increases as
the doping increases and saturates at the optimal doping, then it
decreases slightly in the overdoped regime. While in the even
channel, the resonance frequency increases as the doping decreases,
so that the peak positions of the even and odd channels are closer
and closer as the doping density increases, and the corresponding
resonance frequencies are almost the same in the deeply overdoped
sample. The difference of the intensities in the even and odd
channels increases as the doping density decreases, as seen in
Fig.1(d). The intensity ratio is only $0.1$ in the strongly
underdoped sample, and around 0.4 in the overdoped region. Our
results are qualitatively consistent with the experimental
results~\cite{pailh}. We also examine the relationship between the
resonance frequency and the critical temperature $T_c$ by using an
empirical formula $T_c/T_{c\mathrm{{max}}}=1-c(\delta-0.16)^2$,
where $c=51$ is used to ensure the AF instability to occur at
$\delta=0.02$. The resonance frequency in the odd and even channels
as a function of $T_c$ is plotted in Fig.2 ($\delta\leq0.16$). The
resonance frequency $(\omega^{o}_p)$ is found to be proportional to
$T_c$ in the odd channel, which is in good agreement with the
experimental results~\cite{step,dai,roso}. While in the even
channel, the spin resonance frequency ($\omega^{e}_p$) depends
weakly on $T_c$ in the strongly underdoped sample, being also
qualitatively consistent with the very recent
experiments~\cite{pailh}, in which the resonance frequency in the
even channel is observed to increase in the overdoped sample and to
saturate to a constant in the underdoped sample.

We now address the dependence of the intensity difference of the
spin resonance between the two channels on the parameters $t_\perp$
and $J_\perp$. The intensity ratio as a function of the doping for
different $J_\perp$ is plotted in Fig.3(a). As seen, when $J_\perp$
increases, the intensity ratio decreases, indicating that the
interlayer exchange coupling can strongly affect the even and odd
channels and enhance the difference. However, the intensity ratio
decreases as the doping decreases for all $J_\perp$ we considered,
indicating that our results presented above are robust against the
variation of the interlayer exchange coupling $J_\perp$. On the
other hand, we can also see from Fig.3(a) that even if $J_\perp=0$,
the intensity in the odd channel is still significantly stronger
than that in the even channel, suggesting that the interlayer
exchange coupling is not the only contribution for the difference
between the two channels. In fact, the other contribution comes from
the interlayer single-particle hopping. To show this, we plot the
ratio $\eta$ versus the doping for different hopping constants
$t_\perp$ in Fig.3(b). As $t_\perp$ increases, the ratio decreases.
So, the interlayer single particle hopping also contributes to
enhance the difference between the two channels. Let us also
consider the cases that $t_\perp$ depends on the doping density
$(t_\perp\propto\delta)$ and momentum $[t_\perp\propto (\cos
k_x-\cos k_y)^2]$. Note that, the bilayer splitting is found to be
momentum dependence in Bi-2212 bilayer systems~\cite{feng}, without
observing the bilayer splitting in the nodal direction. We also
examined that our results are robust for different types of the
interlayer hopping constant $t_\perp$, as shown in Fig.3(b).

\begin{figure}

\centering

\includegraphics[scale=0.6]{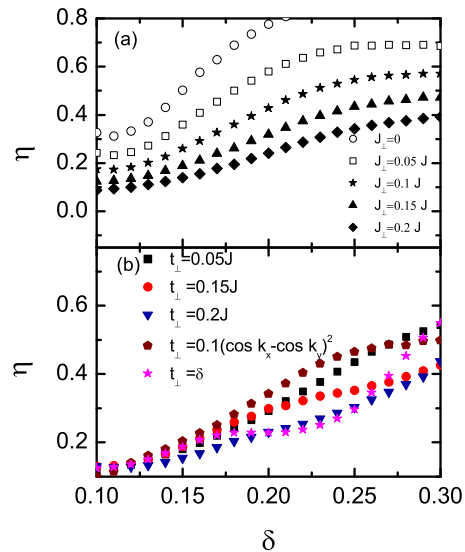}
\caption{(Color on line) (a) The intensity ratio of the spin
resonance between the even and odd channels for different interlayer
exchange coupling with $t_\perp=0.1J$. (b) The intensity ratio of
the spin resonance between the even and odd channel for different
interlayer hopping constants with $J_\perp=0.15J$. }\label{fig3}
\end{figure}

\begin{figure}

\centering

\includegraphics[scale=0.5]{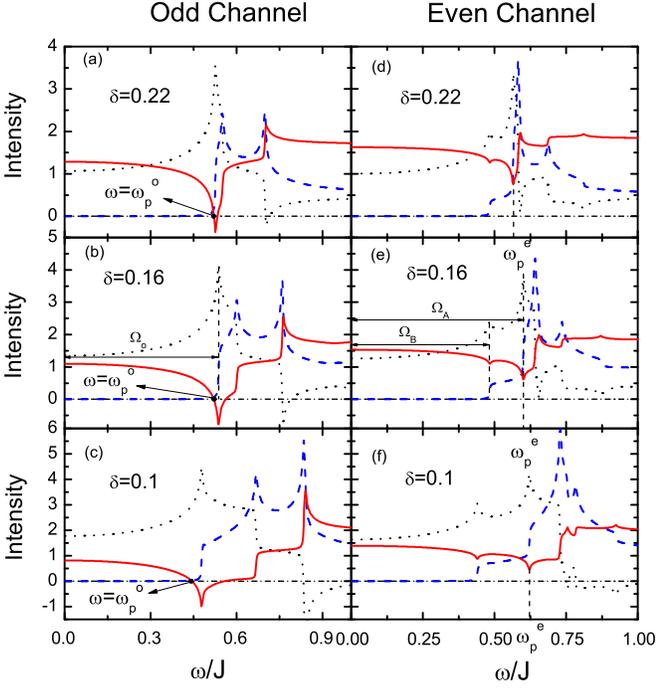}
\caption{(Color on line) The bare spin susceptibilities versus the
frequency $\omega$ at different doping densities in the odd and even
channels with the quasiparticle damping $\Gamma=0.002$. The dotted
lines denote the real parts and the dashed lines imaginary parts,
respectively. The solid lines are the real parts of the RPA factor
$[1+(\widetilde{\alpha} J_{\bf Q}\pm J_\perp)$Re$\chi^{e(o)}_0({\bf
Q},\omega)/2]$ (scaled$\times2$) with $\pm$ signs for the even and
odd channels, respectively. }\label{fig4}
\end{figure}

Now we elaborate the origin of the above features based on the Fermi
surface topology. In Fig.4~\cite{note}, we plot the imaginary and
real parts of the bare spin susceptibility $\chi^{o(e)}_{0}$ in the
odd and even channels with different doping densities. We first
address the spin excitation in the odd channel. As shown in
Fig.4(a-c), the imaginary part of the bare spin susceptibility
approaches to zero as the frequency is below the spin gap. At the
edge of the spin gap, it has a step-like rise which arises from the
flat band near $(\pi,0)$ (Van Hove singularity). As a result, the
real part of the bare spin susceptibility Re$\chi^{o}_0$ develops a
sharp structure near the edge. Consequently, a pole occurs when the
real part of the RPA factor $1+(\widetilde{\alpha} J_{\bf
Q}-J_\perp)$Re$\chi^{o}_0/2$ is equal to zero at the frequency
$\omega^{o}_p$ (slightly below the gap edge) and in the meantime the
imaginary part of the spin susceptibility at $\omega^{o}_p$
approaches to zero due to the spin gap.
This suggests the formation of a spin collective mode, which is
ascribed to be the spin resonance. We can also see from Fig.4(a-c)
that the frequency, at which a step-like rise occurs, increases as
the doping decreases starting from the optimal doping, but it
decrease slightly in the overdoped regime. This explains the doping
dependence of the resonance frequency as presented in Fig.1(c). On
the other hand, the real part of the bare spin susceptibility also
increases with the decrease of doping. This leads to  the pole
position to be more and more below the spin gap edge (due to the
finite damping $\Gamma$, the imaginary part of $\chi^{o}_{0}$ is not
zero slightly below the spin gap) and consequently to an increase in
the renormalized spin susceptibility. For the even channel case, an
obvious difference is seen from Fig.4(d-f) that there are two
step-like rises, instead of one in the odd channel. These two
step-like rises come from the particle-hole scattering in the $B$
and  $A$ bands, respectively, because the spin susceptibility in the
even channel is contributed by the intra-band $A\rightarrow A$ and
$B\rightarrow B$ scatterings as shown clearly in Eq.(4). The
scattering within the $B$ band leads to the step-like rise in
$\chi^{e}_{0}$ at a lower frequency $\Omega_B$, while that in the
$A$ band leads to a higher frequency rise at $\Omega_A$. Since the
rise is larger at $\Omega_A$, the corresponding enhancement of the
real part of $\chi^{e}_{0}$ is larger there. Thus, the spin
resonance peak occurs near $\Omega_A$ in this case. We note that,
due to a smaller vertex $\widetilde{\alpha} J_{\bf Q}+J_\perp$ in
this channel, the pole condition $1+(\widetilde{\alpha} J_{\bf
Q}+J_\perp)$Re$\chi^{e}_0/2=0$ could not be satisfied at a large
doping range. In the meantime, the corresponding imaginary part of
$\chi^{e}_{0}$ is of appreciable value because of the scattering in
the B band. So, the resonance peak in this case is basically a
quasi-resonance peak, with its intensity being much lower than that
in the odd channel as shown in Fig.1(b). Also in contrast to the
case of the odd channel, the frequency, at which the high-energy
step-like rise occurs, increases with the decrease of doping.
Therefore, the spin resonance frequency at the even channel
increases upon reducing the doping as shown in Fig.1(c). On the
other hand, one can see from Fig.4(d-f) that both the real and
imaginary parts of $\chi^{e}_0$ in the lower side of $\Omega_A$ do
not change much with doping. Considering the appreciable increase of
the resonance intensity with the decrease of doping in the odd
channel, it is expected that the difference between  the spin
resonance peak intensities of the two channels increases as the
doping decreases.

\begin{figure}

\centering

\includegraphics[scale=0.6]{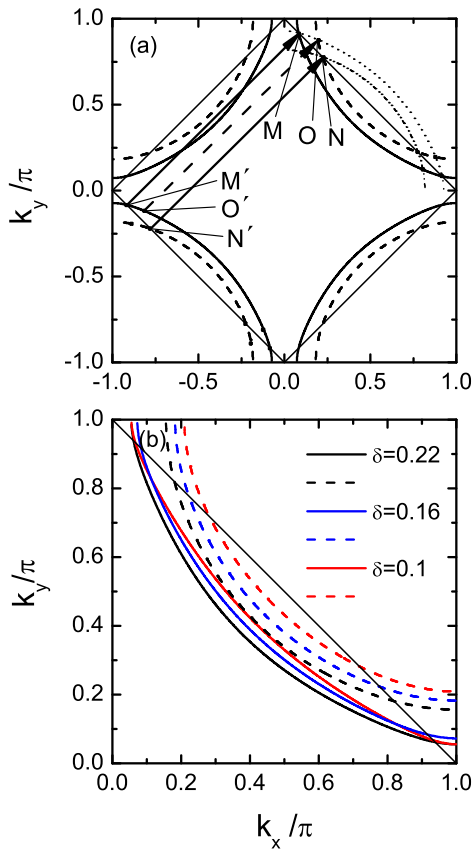}
\caption{(Color on line) (a) The normal state Fermi surface at the
doping $\delta=0.16$ with $t_\perp=0.1J$. The bold solid and dashed
lines are the Fermi surface of the A and B bands, respectively. The
dotted and the dash-dotted lines in the first quadrant are the
$(\pi,\pi)$ shifted images of the A and B band Fermi surfaces in the
third quadrant. The solid and dashed arrows denote the intraband and
interband scatterings, respectively. (b) The Fermi surfaces of the A
(solid lines) and B bands (dashed lines) in the first quadrant of
the Brillouin zone at different doping densities. }\label{fig5}
\end{figure}

These features can be traced to the evolution of the Fermi surface
with doping. We present the normal state Fermi surface in Fig.5. As
discussed above, the step-like rise is near the spin gap edge. In
the zero temperature limit, the bare spin susceptibility [Eq.(3)]
can be rewritten as Im$\chi^{(\alpha,\alpha^{\prime})}({\bf
q},\omega)\propto \sum_{\bf
k}\delta(\omega-\Omega^{(\alpha,\alpha^{\prime})}(\bf k,\bf q))$.
Here, $\Omega^{(\alpha,\alpha^{\prime})}(\bf k,\bf
q)=E^{(\alpha)}_{\bf k}+E^{(\alpha^{\prime})}_{{\bf k}+{\bf q}}$
denotes the energy to break a pair and excite two quasiparticles
from the superconducting condensed state, and  has a minimum of the
exciting energy (MIN$_{\bf k}[\Omega^{(\alpha,\alpha^{\prime})}({\bf
k},{\bf q})]$)  when the wave vector $\bf q$ is at ${\bf
Q}=(\pi,\pi)$ where the spin resonance is observed, which is just
the spin gap. Because of the $d$-wave symmetry of the
superconducting gap and energy band structure, the excitation within
the $A$ band with the minimum excitation energy is the
$M^{\prime}$-to-$M$ excitation as shown in Fig.5(a), while that
within the $B$ band and that of the interband  correspond
respectively to the $N^{\prime}$-to-$N$  and  $O^{\prime}$-to-$O$
excitations, where $M,M^{\prime},N,N^{\prime}$ are the crossing
points (hot spot) of the Fermi surface with the magnetic Brillouin
zone boundary, and $O(O^{\prime})$ are the crossing points of the
$B(A)$ band Fermi surface with the $(\pi,\pi)$ shifted images of the
$A(B)$ band Fermi surface. From Fig.5(b), we can see that the hot
spot of the $B$ band moves towards the nodal direction as the doping
decreases, and consequently the magnitude of the corresponding
superconducting gap decreases. The $A$ band depends weakly on the
doping density, but the magnitude of the superconducting gap
increases with doping, as calculated from the mean-field theory here
and observed in experiments~\cite{shen}. Because the spin excitation
in the odd channel comes from the $O^{\prime}$-to-$O$ excitation,
its spin gap decreases with the decrease of doping. While, the
high-energy step-like rise in the even channel is contributed by the
excitation from $M^{\prime}$ to $M$, it increases with the decrease
of doping.

\section{Summary}
In summary, we have examined the doping evolution of the spin
susceptibility in the even and odd channels in the bilayer
high-$T_c$ superconducting materials based on the bilayer $t-J$ type
model. In the bonding and antibonding band representation, there
exist two channels of the spin excitation according to the intraband
scattering and interband scattering. Each channel has its distinct
resonant mode. In the odd channel, i.e., the interband scattering,
the spin susceptibility shows a strong doping dependence. As the
doping decreases, the intensity increases dramatically and the
resonance frequency is linear with $T_c$. The resonance frequency in
the even channel approaches to that in the odd channel and the ratio
between the two channels is around 0.4 in the overdoped region. As
the doping decreases, the resonance frequency increases and
saturates at the strongly underdoped sample. In addition, it has
been found that the differences of the resonance positions and
intensities between the two channels are enlarged as the doping
decreases. Our results are well consistent with the experiments. We
have elaborated the results based on the topology of the Fermi
surface and the $d$-wave superconductivity.

\begin{acknowledgments}
The work was supported by the RGC grants of Hong Kong(HKU 7050/03P
and HKU-3/05C), the NSFC (10174019,10334090,10429401 and 10525415),
and the 
973 project under the Grant No.2006CB601002.
\end{acknowledgments}


\begin{thebibliography}{99}
\bibitem{note} Here we use a much smaller quasiparticle damping
$\Gamma$ to see the characteristic of the bare spin susceptibility
more clearly. The main results will not change qualitatively as the
quasiparticle damping changes.
\end{thebibliography}
\end{document}